# Single-electron Nano-chip Free-electron Laser

Yen-Chieh Huang,[1,a)] Luo-Hao Peng,[1] Hossein Shirvani,[1] Wen-Chi Chen,[1] Karthickraj Muthuramalingam,[2] and Wei-Chih Wang[2,3,4,5]

**AFFILIATIONS**

[1] Institute of Photonics Technologies, National Tsing Hua University, Hsinchu 30013, Taiwan

[2] Institute of Nanoengineering and MicroSystems, National Tsing Hua University, Hsinchu 30013, Taiwan

[3] Power Mechanical Engineering, National Tsing Hua University, Hsinchu 30013, Taiwan

[4] Department of Electrical Engineering, University of Washington, Seattle 98195, Washington, USA

[5] Department of Mechanical Engineering, University of Washington, Seattle 98195, Washington, USA

[a)] Authors to whom correspondence should be addressed: ychuang@ee.nthu.edu.tw

**ABSTRACT**

A conventional free-electron laser is useful but large, driven by a beam with many relativistic electrons. Although, recently, keV electron beams have been used to excite broadband radiation from material chips, there remains a quest for a chip-size free-electron laser capable of emitting coherent radiation. Unfortunately, those keV emitters from electron microscopes or dielectric laser accelerator usually deliver a small current with discrete moving electrons separated by a distance of a few or tens of microns. To envisage a chip-size free-electron laser as a powerful research tool, we study in this paper achievable laser radiation from a single electron and an array of single electrons atop a nano-grating dielectric waveguide. In our study, thanks to the strong coupling between the electron and the guided wave in a structure with distributed feedback, a single 50-keV electron generates 1.5-μm laser-like radiation at the Bragg resonance of a 31-μm long silicon grating with a 400-nm thickness and 310-nm period. When driven by a train of single electrons repeating at 0.1 PHz, the nano-grating waveguide emits a strong laser radiation at the second harmonic of the excitation frequency. A discrete spectrum of Smith-Purcell radiation mediated by the waveguide modes is also predicted in theory and observed from simulation in the vacuum space above the grating waveguide. This study opens up the opportunity for applications requiring combined advantages from compact high-brightness electron and photon sources.

## I. INTRODUCTION

A free-electron laser[1] (FEL) generates laser radiation from electrons propagating in vacuum with magnetic fields or material structures. To achieve lasing, the injected electrons experience radiation feedbacks in the laser structure and are collectively bunched into the radiation cycles to generate stimulated emission of radiation[2]. For relativistic electrons, a magnetic undulator[3] is often used to induce transverse motion of the electrons and couple the electron energy to the amplification of the radiation field. Some vacuum electronic devices, such as backward wave oscillators[4] and Cherenkov lasers[5], usually adopt a slow-wave structure to match the longitudinal velocities between sub-relativistic electrons and a radiation wave for continuous energy transfer. A Smith-Purcell radiator[6], driven by an electron beam above a metal grating, can also generate broadband radiations at different directions above the grating. Additional resonances, such as cavity feedback, are needed to generate narrow-line stimulated Smith-Purcell radiation[7].

Recently, research on laser-driven chip-size accelerators, known as dielectric laser accelerators or DLAs[8,9], and their keV injectors[10] have attracted wide attentions and inspired new opportunities for applications utilizing compact electron sources. By running the acceleration process in reverse, an accelerator chip could function as a radiation chip. It is possible that a monolithically integrated accelerator and radiator on a chip[11] could be realized with good efficiency in the near future. Currently, high-brightness keV electrons are already available from an electron microscope. An electron microscope equipped with a built-in FEL chip can be a powerful pump-probe tool for material research. However, a tiny electron emitter can only generate a small electron current. For instance, the average temporal separation of two adjacent electrons in a nano-ampere current from a DLA or a transmission electron microscope (TEM) is about ~100 ps. At 50 keV, the spatial separation of the two electrons is about a centimeter. If the radiation chip has a longitudinal length of the order of a millimeter, there is only one electron driving the chip at a time. To generate laser-like radiation in one electron transit, it requires strong coupling between the electron and the radiation mode. In this scenario, stimulated emission from many gradually bunched electrons in a conventional FEL is irrelevant to the radiation process.

In the optical frequencies, metal is lossy, making dielectric the choice for most optical components. A conventional Fabry-Perot resonator with a many-wavelength length is too long to provide any optical feedback to an electron in one transit. However, in a dielectric grating with nano-periodicity, an electron with an extended Coulomb field could resonantly excite and amplify the distributed optical feedback from individual grating grooves with little time delay. Although Cherenkov radiation in a photonic crystal[12,13] or metamaterial[14], and Smith-Purcell radiation from metallic nano-grating[15,16] or dielectric-metal hybrid structures[17,18] have been studied in the past, this study utilizes a much simpler dielectric-grating waveguide to maximize the electron-wave coupling and build up narrow-line radiation in a single electron transit. The highly directional and focused wave from the waveguide output can be a major advantage for a downstream application.

In the following, we first introduce a few theoretical guidelines for designing the proposed nano-chip FEL, and then employ mode-expansion theory and simulation code to confirm the parameters for a realistic design. Finally, we perform time-domain simulations to understand the radiation mechanism and device performance for a single-electron excited nano-grating waveguide. Before we conclude this paper, we demonstrate coherent harmonic generations from the nano-grating waveguide by driving it with a periodic pulse train of single electrons from a DLA operating at a sub-harmonic frequency. Finally, we summarize the study of this paper in the last section.

## II. BASIC DESIGN THEORY

Figure 1 depicts the proposed dielectric-grating waveguide, in which an electron propagates a distance $l_{ip}$, called the impact parameter, above the grating surface. The structure is a corrugated dielectric film on a dielectric substrate. Although the substrate is not essential for wave guiding, a thick enough substrate is often necessary to support a sub-micron thick laser waveguide. Without loss of generality, the surface corrugation is assumed to have a rectangular shape with a period of $\Lambda_g$ and depth of $t_g$. The smooth film layer under the grating has a thickness of $t_f$. In this work, we assume the grating and the film layer are the same optical material. To guide the radiation, the grating waveguide has a refractive index $n_f$, which is larger than that of substrate $n_s$. In the vacuum region, there will be Smith-Purcell radiation, which will be shown below as correlated to the radiation inside the waveguide. The two-dimensional (2-D) structure in Fig. 1(a) is first adopted for our theoretical analysis. To plan a real experiment, we perform numerical simulations to further study the finite-width three-dimensional (3-D) structure in Fig 1(b). Figure 1(c) shows the scanning-electron-microcopy images of a fabricated silicon grating with 300-nm period and 4-μm width for an on-going feasibility study. The radiation wave is guided in the corrugated dielectric film above a substrate. On the grating surface, the electron's kinetic energy is transferred to the evanescent field of the transverse-magnetic (TM) waveguide mode, when the electron velocity matches the mode field's phase velocity. Before presenting a more detailed numerical study, we first list below a few physics laws to estimate the parameters to generate a radiation with a desired wavelength.

For the radiation to be guided in the film layer, the incidence angle $\theta$ of the guided wave at the film and substrate interface must be larger than the critical angle of total internal reflection $\theta_c$ or

$$\sin\theta > \sin\theta_c = \frac{n_s}{n_f}. \tag{1}$$

For a silicon grating of $n_f$ = 3.4 on a glass substrate of $n_s$ = 1.5, the condition for total internal reflection is $\theta$ > 26°. Assuming the surface corrugation is a small perturbation to the radiation modes in the film layer, the velocity matching to ensure continuous energy transfer from the electron to the radiation field is primarily governed the Cherenkov condition, given by

$$\cos\phi = \frac{1}{n_f \beta_e}, \tag{2}$$

where $\beta_e = v_e/c_0$ is the electron velocity $v_e$ normalized to the vacuum speed of light $c_0$ and $\phi = 90° - \theta$ is the Cherenkov angle. From (1) and (2), one obtains the range of the speed of the electron for a guided Cherenkov radiation in the film layer,

$$\frac{1}{n_f} < \beta_e < \frac{1}{n_s}. \tag{3}$$

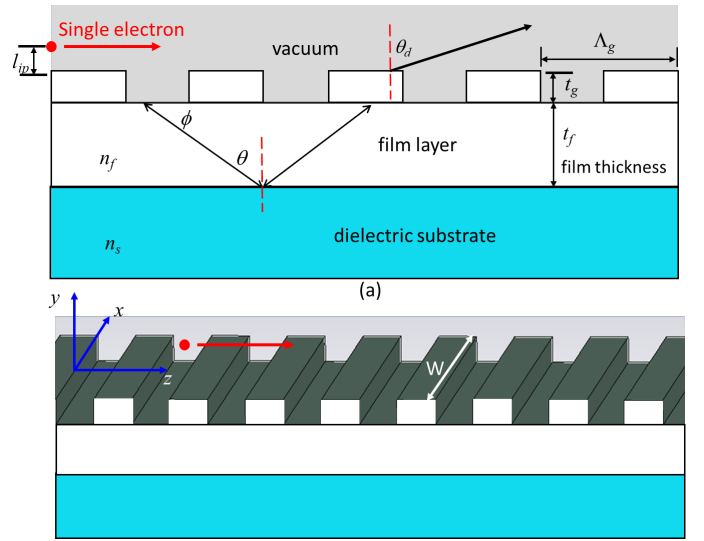

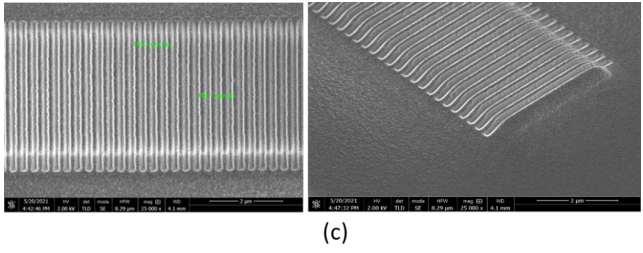

(c)

**FIG. 1.** A proposed dielectric-grating waveguide as a nano-chip FEL driven by a single keV electron. The distributed optical feedback and the guided field in the grating film increase the electron-wave coupling for the radiation process. (a) A 2-D configuration used in our theoretical modal analysis. (b) A 3-D configuration used for our time-domain radiation simulation. (c) Scanning-electron-microscope images of a fabricated silicon grating with 300-nm period and 4 μm width for an on-going feasibility study.

The condition $\beta_e > 1/n_f$ is the Cherenkov threshold in a bulk dielectric of refractive index = $n_f$. For a silicon film ($n_f$ = 3.4) on a glass substrate ($n_s$ = 1.5), the electron-energy range to generate the guided radiation, according to Eq. (3), is between 24 and 175 keV.

In the vacuum region, the surface field of the guided mode synchronously propagates with the electron along $z$. The synchronous field is evanescent with an exponential decay constant $\alpha$ satisfying the dispersion relationship

$$(\omega/c_0)^2 = (\omega/v_e)^2 - \alpha^2, \quad (4)$$

where $\omega$ is the radiation frequency and $c_0$ is the speed of light in a vacuum. This dispersion relationship defines a mode-field depth above the grating surface, given by

$$h = \frac{1}{\alpha} = \frac{\beta_e \gamma \lambda_0}{2\pi}, \quad (5)$$

where the Lorentz factor of the electron $\gamma$ is about unity for cases subject to (3) and $\lambda_0$ is the vacuum wavelength of the radiation. To have enough coupling between the electron and the mode field, one usually sets an impact parameter $l_{ip}$ ~ $h$. For a keV electron with $\beta_e$ ~ 0.5, the impact parameter $l_{ip}$ is approximately one tenth of the radiation wavelength.

The surface corrugation is meant to provide distributed optical feedback to the resonant mode, so that narrow-band coherent radiation can be established through a single electron transit. In practice, to avoid incident electrons charging the structure, the surface of the grating waveguide can be coated with a metal layer much thinner than the skin depth. Consider each grating tooth along $y$ as a short transmission line terminated with a perfect conductor. The impedance seen by an incidence wave has a period of a half wavelength[19] in $y$. To have the maximum impedance contrast for the grating[20], the depth of the corrugation, $t_g$, can be a quarter wavelength of the mode field in $y$, given by

$$t_g = \frac{\lambda_y}{4} = \frac{\lambda_0}{4 n_f \sin\phi} = \frac{\beta_e \lambda_0}{4\sqrt{\beta_e^2 n_f^2 - 1}}, \quad (6)$$

where the Cherenkov condition, Eq. (2), has been used to derive the expression as a function of the normalized electron velocity $\beta_e$.

Given a Cherenkov angle $\phi$, a thick film $t_f$ could contain high-order transverse modes with slow group velocities and weak fields. It is therefore desirable to excite a fundamental mode in the dielectric film. To have a single-mode waveguide at a design wavelength $\lambda_0$, the thickness of the waveguide film must satisfy the condition[21]

$$t_f < \frac{\lambda_0}{2\sqrt{n_f^2 - n_s^2}}. \quad (7)$$

Supposing the design wavelength is 1.5 μm, the single-mode film thickness is $t_f$ < 246 nm for $n_f$ = 3.4 (silicon film) and $n_s$ = 1.5 (glass substrate).

A grating structure can provide two types of distributed-feedback resonance to an electromagnetic wave. The first type is Bragg resonance[22], and the second one is backward-wave oscillation[23]. The Bragg resonance establishes a standing wave with two counter-propagating components in the grating waveguide. For instance, highly stable and useful single-frequency distributed-feedback diode lasers[24] are based on Bragg resonance in a semiconductor gain waveguide. In a backward-wave oscillator, a backward-wave mode has a group velocity in the opposite direction of the electron propagation. With single-electron excitation, it is likely that the co-propagating component of the Bragg mode will have a lower lasing threshold, although gain competition from the backward-wave mode is possible. As a first-order design for the nano-chip FEL, we aim to build up the low-threshold Bragg mode.

The first-order Bragg resonance in the grating is given by the condition,

$$2k_z = k_g = \frac{2\pi}{\Lambda_g} \quad (8)$$

where $k_z = (k_0 n_f) \times \cos\phi = 2\pi n_f / \lambda_0 \times \cos\phi$ is the propagation constant of the wave along the electron axis $z$. Physically, it means the roundtrip reflection phase of the electromagnetic field over a grating period is $2\pi$. This condition allows each period of the grating to form a small resonator with a half-wavelength length. From Eqs. (2) and (8), the grating period depends on the speed of the electron $\beta_e$, given by

$$\Lambda_g = \beta_e \frac{\lambda_0}{2}. \quad (9)$$

It is straightforward to show that the Bragg condition along the axial direction is the same as the Littrow-grating diffraction condition[25], for which diffraction angle of an incident wave is the same as the incident angle of the wave. This is illustrated in Fig. 1(a), wherein each bouncing ray has two arrows to denote counter-propagating zigzag components oscillating inside the structure.

Equations (3, 5-7, 9) offer a set of theoretical guidelines to perform the first-order design for a nano-chip FEL structure prior to numerical optimizations. We begin by supposing that the available electron energy is 50 keV, corresponding to an electron speed of $\beta_e = 0.41$. We further assume a target laser wavelength of $\lambda_0 = 1.5$ μm for a silicon-on-glass grating waveguide. From Eq. (5), the impact parameter of the electron is calculated to be 98 nm. The quarter-wave grating depth, $t_g$, calculated from Eq. (6), is 160 nm. To excite single-mode radiation at 1.5 μm in the film layer, Eq. (7) gives a maximum film thickness of $t_f = 246$ nm. From Eq. (9), the grating period matched to the Bragg resonance is 308 nm. Note that the chosen film thickness is only 1.5 times the groove depth of the grating, which challenges the perturbation assumption made for the first-order design. However, increasing the waveguide thickness $t_f$ could weaken the coupling between the electron and the mode field. For what follows, we round those estimated parameters and use them for more detailed numerical analysis. Table 1 lists the chosen parameters used for our numerical studies for a nano-chip FEL chip emitting at 1.5 μm.

**TABLE I.** The first-order design parameters for a 1.5-μm nano-chip FEL with a silicon ($n_f = 3.4$) grating waveguide on a glass substrate ($n_s = 1.5$).

| design wavelength (μm) | electron energy (keV) | grating period $\Lambda_g$ (nm) | grating depth $t_g$ (nm) | film thickness $t_f$ (nm) | impact parameter $l_{ip}$ (nm) |
|---|---|---|---|---|---|
| 1.5 | 50 | 310 | 160 | 240 | 100 |

## III. MODE ANALYSIS

The dispersion of the grating waveguide determines the radiation frequency subject to the velocity matching between the electron and the guided mode field. To find the dispersion of a periodic structure, most theories consider steady-state loss-free eigenmodes in an infinitely long periodic waveguide satisfying the Floquet theory[26]. To account for the transient radiation excited by a single electron, we adopt the mode-expansion formulism developed by Peng, Tamir, and Bertoni[27] (the PTB model), in which both steady-state and leaky modes are included without assuming the grating grooves providing a small perturbation in a 2-D dielectric waveguide. With the design parameters in Table 1, Fig. 2(a) shows the dispersion curves calculated from the PTB model for the TM modes in the proposed grating-waveguide structure. The guided-mode curves are in the region between the light lines of the film and the substrate. The Bragg resonances, marked with colored squares, are located at the intercepting points of the dispersion curves and the vertical line $k_z/k_g = 0.5$. It is seen that the fundamental Bragg mode has a resonant frequency at 0.2009 PHz or a vacuum wavelength of almost 1.5 μm, which is very close to the design value of 1.5 μm. The first Bragg point is intercepted by the electron line with a slope associated with 50.79-keV energy, which is just 1.8% higher than the design value of 50 keV. All the steady-state mode curves have a zero slope at the Bragg resonances, where the group velocity of a resonating standing wave is zero. The red-dashed lines are the dispersion curves of the low-loss modes found in the PTB model. However, the imaginary part of their $k_z$ is only $10^{-10} \sim 10^{-18}$ of the real part. Some branch of them has a negative slope. For instance, the 50.79-keV electron line intercepts the red-dashed curves at frequencies 0.2028 PHz and 0.2463 PHz, capable of exciting backward radiations inside the grating waveguide.

As a comparison, Fig. 2(b) plots the dispersion curves of the TM modes calculated by the simulation code COMSOL[28]. Figure 2(a and b) show that the first 3 Bragg frequencies agree with each other within 1-2%. In (b), the first Bragg mode is intercepted by a 50.73-keV electron line, which matches very well to the 50.79 electron line in (a). The insets in Fig. 2(b) show the TM-mode-field ($H_x$) patterns at the Bragg resonances, indicating strong wave guiding in the silicon-grating region. Unlike the PTB model, COMSOL does not provide solutions of those low-loss mode curves shown as red dashed curves in Fig 2(a).

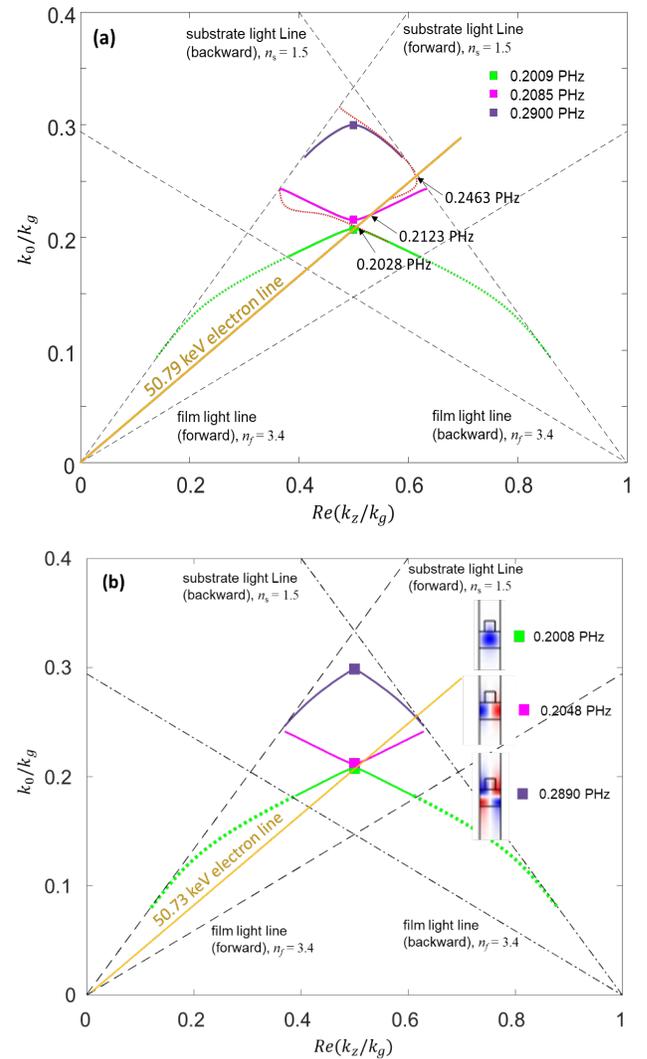

**FIG. 2.** The dispersion diagram calculated by (a) the PTB model and (b) COMSOL for the TM modes in the dielectric-grating waveguide specified in Table 1. Both plots show the first Bragg resonance at 1.5 $\mu m$ (0.200 PHz), intercepted by a ~50 keV electron line. In (a), the red dashed curves are low-loss leaky modes predicted by the PTB model. In (b), the zero-loss mode curves and the electron line match very well to those calculated by the PTB model. The insets show the $H_x$ field patterns at the Bragg resonances, indicating strong guiding in the grating region. Outside the region boxed by the light lines, the mode is lossy and is not guided.

## IV. TIME-DOMAIN SIMULATION FOR RADIATION GENERATION

In practice, a grating structure has a finite length and width, as shown in Fig. 1(b). The calculations in Fig. 2 only consider an infinitely long 2-D periodic structure without a width along $x$. In the following, we use the 3-D time-domain simulation code, CST Studio Suite[29], to study a practical case for the proposed nano-chip FEL. To compare with the previous 2-D calculations, we set a grating width of 4 μm in the range of $x = \pm 2$ μm, so that the width is about 10 times the 1.5-μm wavelength for the first Bragg resonant mode. Therefore, the reflection feedback from the $x$ boundaries cannot reach a single electron traversing along $z$. The 3-D structure consists of 100 grating periods along $z$, having a total length of 31 μm. The rest of the design parameters are listed in Table I, except that we compare below the radiation characteristics for a waveguide grating with $t_f$ = 240 nm and a bulk grating with $t_f$ = ∞. The transit time for a 50-keV electron along the 31-μm long grating is a quarter of a picosecond. For a beam current less than 0.6 μA, on average, there is at most one electron exciting the structure at a time.

Figure 3 shows the calculated TM-field ($H_x$) patterns on the $y$-$z$ plane (cut at $x = 0$) for the grating structures with (a) $t_f$ = ∞ and (b) $t_f$ = 240 nm at 0.2 and 0.6 ps after the electron is injected from the left edge of the structure. The colored dots are the locations of the field probes installed in the simulation for presenting the signals in Fig. 4. The structure in (a) is a bulk grating with no wave guiding in the dielectric layer; whereas the structure in (b) is a grating waveguide confining the radiation in the film layer. It is seen from (a-1) that, at 0.2 ps, a Cherenkov radiation cone following the electron is extended into the whole dielectric region under the grating; at the same time, Smith-Purcell radiation appears in both the vacuum and dielectric regions. In (a-2), radiations dissipate into the whole space after the electron leaves the grating. In (b-1), a single electron excites strongly confined radiation inside the grating waveguide. The enlarged field pattern in the inset clearly shows the characteristic Bragg resonance with $\Lambda_g = \lambda_z/2$, where $\lambda_z$ is the longitudinal wavelength of the radiation mode. After the electron exits the structure, the radiation field stored inside the waveguide starts to ring down over time. Figure 3(b-2) shows the field patterns recorded at 0.6 ps, indicating emission of quasi-coherent radiation with well-defined wavefronts in both the forward and backward directions. The inset is the ring-down field at the downstream output of the waveguide, which starts to emit at 0.25 ps when the electron just exits the structure and rolls off over a period of about 1 ps.

Figure 4 shows the Fourier spectra of the TM radiation field, $H_x(f)$, detected by the field probes installed at the colored dots for (a) the bulk grating and (b) the waveguide grating. On the $y$-$z$ plane, the (0, 0, 0) origin of the coordinate system is donated as O in the insets, located at the top left edge of the fist grating tooth. A 50-keV electron is injected along $z$ at the coordinates (0, 0.1, 0) in units of $\mu m$. For the waveguide grating, the film-glass interface is located at (0, 0, –0.4). Since the structure contains 100 grating periods with 310-nm periodicity, the downstream end of the structure is located at (0, 0, 31). The amplitudes of all the curves are normalized to the peak amplitude of the cyan curve in Fig. 4(a-1), which is the Fourier transform of $H_x$ detected at the downstream output point of the silicon bulk grating (0, –0.28, 31). Also, in (a-1), the orange curve is the signal recorded at (0, –0.28, 15.73) or in the dielectric slightly below the longitudinal center of the bulk grating. Both curves in (a-1) are broadband. The signal below the grating (orange curve) is stronger and modulated with weak resonances from the surface grating. Figure 4(a-2) shows the Smith-Purcell radiation in the vacuum region, detected at the upstream point (0, 7.5, 0) and downstream point (0, 7.5, 31) as only a few percent of the Smith-Purcell radiation immediately below the dielectric grating. Note that while (a-1) and (a-2) have been plotted on the same vertical range, the amplitude of $H_x(f)$ in (a-2) has been multiplied by 10 to enhance visibility of the detected vacuum radiation.

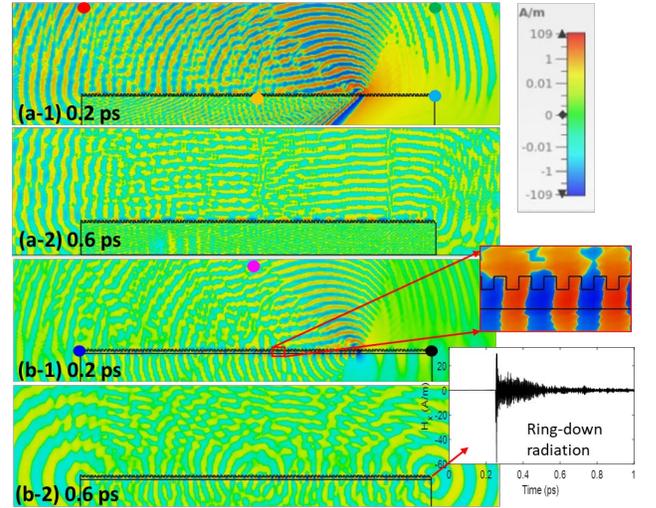

**FIG. 3.** Single-electron excited TM ($H_x$) field patterns calculated by CST for (a) a bulk grating with $t_f$ = ∞ and (b) a waveguide grating with $t_f$ = 240 nm, when the electron is still in the grating at 0.2 ps and has left the grating at 0.6 ps. Colored dots are field probes recording the signals in Fig. 4. In (a), the electron generates scattered Cherenkov radiation and Smith-Purcell radiation. In (b), the electron excites strongly guided radiation and the stored quasi-coherent radiation emits from the waveguide output. Inset for (b-1): Guided Bragg-mode field satisfying the Bragg condition in Eq. (8). Inset for (b-2): Ring-down field from the waveguide output.

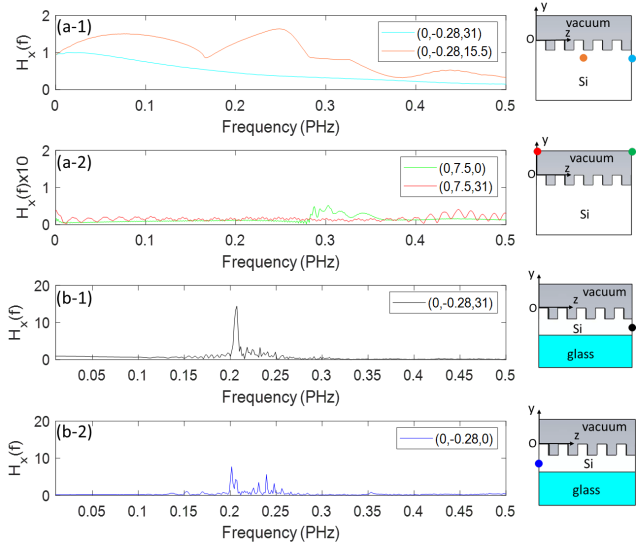

**FIG. 4.** (a) Fourier spectra of $H_x$ (a-1) in the dielectric of the bulk grating and (a-2) in the vacuum above the bulk grating. (b) Fourier spectra of $H_x$ emitted (b-1) in the forward direction and (b-2) in the backward direction of the waveguide grating. Insets illustrate the locations of the field probes (colored dots) recording the $H_x$-field signal. All the amplitudes are normalized to the peak value of the cyan curve. For clarity, the Fourier amplitude in (a-2) is enlarged by 10 times. The spectral amplitude of the forward radiation from the grating waveguide is narrowest and strongest.

In Fig. 4(b-1), the forward radiation at the downstream output point of the waveguide grating, detected at (0, -0.28, 31), has a narrow Bragg resonance at 0.2043 PHz, which matches the theoretical and COMSOL predictions. In Fig. 4(b-2), the backward radiation detected at (0, -0.28, 0) from the waveguide grating is a about 3 times weaker at 0.2 PHz and contains a few satellite peaks at slightly higher frequencies consistent with the additionally marked resonances in Fig. 2(a). Compared with the radiation from the bulk grating driven by a 50-keV electron, the spectral amplitude of the narrow-line radiation emitted from the grating waveguide is more than 10 times higher. A stronger signal in the forward direction also signifies an amplification gain from the co-propagating electron.

When the radiation modes resonantly build up inside the waveguide, some radiation can transmit through the grating and become useful radiations in the vacuum region for applications. Consider the grating formula for a dielectric transmission grating:

$$\frac{\Lambda_g}{\lambda_0} n_f \sin\theta - \frac{\Lambda_g}{\lambda_0} \sin\theta_d = m \quad (10)$$

where $\theta_d$ is the diffraction angle above the grating and $m$ is the diffraction order number. Choosing $\theta = 90° - \phi$ with $\phi$ being the Cherenkov angle and using Eq. (2) in Eq. (10) result in

$$\sin\theta_d = \frac{1}{\beta_e} - m\frac{\lambda_0}{\Lambda_g}, \quad (11)$$

which is simply the Smith-Purcell radiation angle[30] above a grating for a given electron velocity, grating period, and radiation wavelength.

Smith-Purcell radiation above a grating is usually broadband, having different wavelength components emitting along different directions. For the diffraction angle $\theta_d$ to exist in (11), the wavelength $\lambda_0$ must fall into the range

$$\frac{\Lambda_g}{m}(\frac{1}{\beta_e} - 1) < \lambda_0 < \frac{\Lambda_g}{m}(\frac{1}{\beta_e} + 1) \quad . \quad (12)$$

For $\beta_e$ = 0.41 (50 keV) and $\Lambda_g$ = 310 nm, the wavelength ranges of the Smith-Purcell radiation described by Eq. (2) are 446 nm < $\lambda_0$ < 1066 nm (0.28 ~ 0.67 PHz) and 223 nm < $\lambda_0$ < 533 nm (0.56 ~ 1.34 PHz) for $m$ = 1 and 2, respectively. Since the narrow-band radiations of the waveguide modes could also emerge as the Smith-Purcell radiation, one would expect a broad radiation spectrum with sharp peaks in the vacuum region. Indeed, Fig. 5 shows a few narrow lines in the $H_x$ spectrum detected at the longitudinal center above the grating waveguide (probe coordinates = (0, 7.5, 15.5)). Again, the amplitude of $H_x(f)$ in the plot is normalized to the peak value of the cyan curve in Fig. 3(a). When compared with the TM-field spectrum above the bulk grating in Fig. 3(b), the radiation in Fig. 4 is more intense and spikier, containing a strong peak from the leaked Bragg mode at 0.2 PHz.

The quasi-coherent radiation emitted from the waveguide ends have the highest spectral brightness. To estimate the efficiency of the useful radiation from the waveguide, we extract the field data versus time from the CST simulation and integrate its power density over a ring-down time of 1 ps across the waveguide aperture 0.24 μm × 6 μm. The radiation energy in the forward direction is about 2.7 atto-joules and that in the backward direction is about 0.51 atto-joules. The much higher forward radiation energy again implies an amplification gain for the synchronous radiation field co-propagating with the moving electron. Given the injection energy of 8 femto-joules for a 50-keV electron, the conversion efficiency for the laser-like energy is about 4×10$^{-4}$.

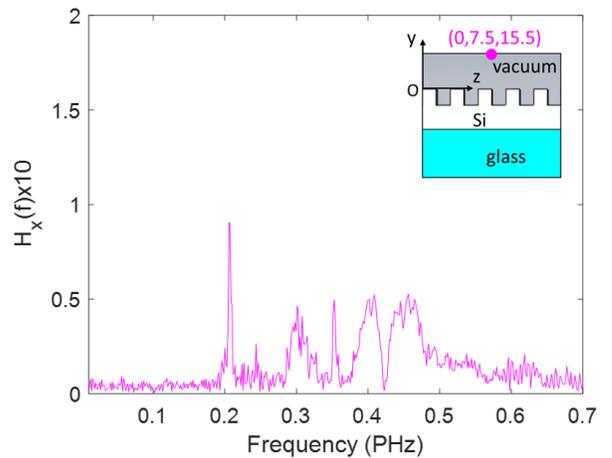

**FIG. 5.** The normalized $H_x$-field spectrum of the Smith-Purcell radiation detected at (0, 7.5, 15.5) μm above the waveguide grating. When compared with that in Fig. 3(a-2), this spectral signal is 3-10 times stronger and contains mode lines emitted from the waveguide.

## IV. HARMONIC GENERATION

In radiation generation, when the electron bunch length is significantly shorter than the radiation wavelength, all the radiation fields of the electrons can add up coherently and the radiation power is proportional to the square of the number of electrons in the bunch. Such intense radiation is dubbed as electron superradiance[31,32]. Furthermore, if the electron bunches repeat at a sub-harmonic of the radiation frequency, the spectral brightness at the radiation frequency is enhanced due to the constructive interference from the periodic electron bunches. The envisaged DLA is to produce a periodic electron pulse train at optical frequencies. Assume that a 50-keV DLA generates a beam with one electron in each optical cycle and the optical cycle repeats at 0.1 PHz (a DLA driven by a 3-μm laser). By using the CST simulation code, we study the radiation of such an electron train injected into the proposed nano-chip FEL with the first Bragg resonance at 0.2 PHz. The design parameters for the structure are the same as those listed in Table 1.

In the CST model, we individually inject 25 electrons at 0.1 PHz into the structure with an impact parameter of 100 nm. The first electron is injected along z at t = 0 at the coordinates (0, 0.1, 0). The last electron exits the downstream end of the grating at 0.5 ps at the coordinates (0, 0.1, 31). Figure 6(a) shows the field pattern of $H_x$ recorded at 0.2 ps, wherein the periodic array of the 0.1-PHz electrons is seen to occupy roughly the first three-quarter of the grating section and generates a guided field pattern with a period of $4\Lambda_g$ (inset). However, the low-threshold Bragg mode at 0.2 PHz can be resonantly built up from the coherent excitation of the 0.1-PHz electron train[33]. As seen from the $H_x$-field pattern of the same area recorded at 0.9 ps in Fig. 6(b), both ends of the waveguide emit coherent radiation with well-defined wavefronts. The inset shows that the field pattern inside the film layer has a period of $\lambda_z = 2\Lambda_g$ (inset), as expected from the Bragg condition for the resonance at 0.2 PHz.

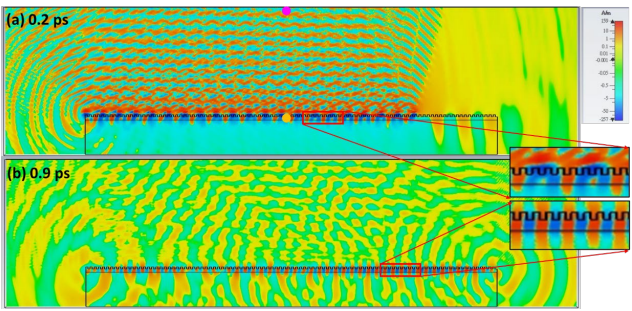

**FIG. 6.** The $H_x$ field patterns excited by a train of 25 electrons repeating at 0.1 PHz. (a) At 0.2 ps, the periodic field of the 50-keV electrons penetrates the film layer to generate a field pattern with a period of $4\Lambda_g$ (inset). (b) At 0.9 ps, after all the electrons exit the structure, both ends of the grating waveguide continue to emit the coherent radiation at the 0.2-PHz Bragg resonance. Insets are the magnified view of the field patterns in the film layer, indicating that the 0.2-PHz Bragg-mode field in (b) has a period half that of the 0.1-PHz electron-excitation field in (a). The colored dots denote the locations of the field probes for the signals to be presented in Fig. 7.

To show the much-increased brightness of radiation from the harmonic generation, Fig. 7 plots the normalized TM-field spectrum, $H_x(f)$, at (a) the downstream output of the waveguide (0, –0.28, 15.5), (b) the upstream output of the waveguide (0, –0.28, 0), and (c) the mid-point vacuum above the grating (0, 7.5, 15.5). As expected from the coherent enhancement of the periodic electron train, the forward radiation signal in (a) has a strong narrow peak at 0.2 PHz and a few small peaks at the harmonics of 0.1 PHz. The peak amplitude at 0.2 PHz is 14 times stronger than that in Fig. 4(b-1) for the same structure excited by a single electron. The backward-radiation spectrum in (b) also shows factor of 20 increased in the peak at 0.2 PHz in a 0.1-PHz frequency comb. The field enhancement factor varies at different locations because the grating waveguide is a resonator with a nonlinear gain. For instance, in our simulation data, the peak spectral amplitude at the waveguide center has reached a value of 665 at 0.2 PHz on the normalized scale. The 0.2-PHz spectral peak of the forward radiation is 1.3 times stronger than that of the backward radiation due to the forward amplification gain from the co-propagating electron train.

Compared with the broadband spectra in Fig. 4(a-2) and Fig. 5, the Smith-Purcell radiation in Fig. 6(c) is fully coherent in the vacuum region, showing a high contrast frequency comb with 0.1-PHz comb spacing. According to Eq. (11), the radiation direction of each comb component depends on the radiation frequency, thus the relative amplitudes of the comb peaks vary with the location of the field probe. However, the overall spectral amplitude of the radiation in vacuum is about two orders of magnitude lower than that emitting from the waveguide outputs.

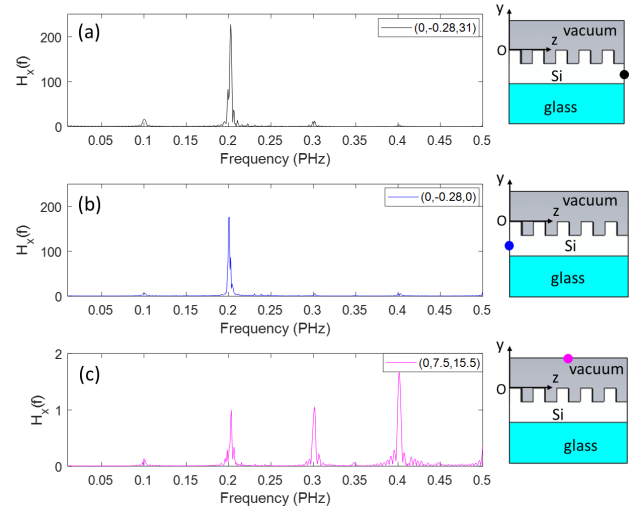

**FIG. 7.** The spectra of the normalized TM field, $H_x$, excited by a train of 25 discrete electrons repeating at 0.1 PHz. Both the (a) forward and (b) backward spectra from the grating waveguide show a much-enhanced Bragg peak at 0.2 PHz in a 0.1-PHz frequency comb. (c) High harmonic radiation escapes from the waveguide and becomes coherent Smith-Purcell radiation in the vacuum above the grating surface.

## V. CONCLUSION

High-brightness electron sources, such as a TEM, usually deliver an average current between nA and μA. It is desirable to use the high-brightness beam to generate high-brightness radiation. However, on average, a nA and µA electron beam can only deliver ~1 electron at a time to drive a radiation emitter with a length of a few tens of microns. We present in this paper achievable narrow-line radiation from a nano-grating waveguide excited by a single keV electron, thanks to the strong energy coupling between the electron and the resonance modes in the grating waveguide.

We describe the excited radiation in the waveguide as Cherenkov radiation resonated by a grating structure. The thickness of the grating waveguide is set to have a single transverse mode at the design wavelength to increase the particle-wave coupling in the surface field. A quarter-wave depth for the grating groove is chosen to maximize the impedance contrast seen by the drive electron. When excited by a single electron with an extended Coulomb field, the distributed feedback from the grating grooves, including the Bragg and backward-wave resonances, helps establish the strength and coherence of the narrow-line radiation. We have developed a set of simple formulas for designing a nano-grating FEL chip. By using a 50-keV electron to drive the chip at the Bragg resonance, the radiation performance at 1.5 µm is consistent with the predictions from the formulas, and from the rigorous calculations by the PTB mode-expansion model, COMSOL, and CST. A single electron is a delta-function excitation source, capable of exciting all possible modes in the dielectric waveguide. The low-loss scattering modes calculated from the PTB model further explain the observation of a few radiation peaks in CST simulations. In our CST simulation, we show the radiation generated from the nano-grating with a guiding structure is narrow-line and >10 time more intense, when compared with that generated from the same nano-grating without a guiding structure.

With the rapid advancement on the development of structure-based laser-driven accelerator, a DLA delivering a keV electron beam is becoming available[34,35]. In our study, we assume that a DLA driven by a laser with a 3-µm wavelength can delivers a 50-keV electron in each optical cycle, repeating at a 0.1 PHz rate. In our CST simulation, we injected 25 such electrons into the nano-grating-waveguide structure to show high-brightness harmonic radiations with the strongest peak at the Bragg resonance of 0.2 PHz. We have also developed a theory to show that the discrete waveguide modes can leak out the waveguide and become the Smith-Purcell radiation in the vacuum above the grating. Our computer simulation confirms the coherent Smith-Purcell radiation mediated by the radiation modes in the waveguide. If a future DLA could fill each optical cycle with $N$ electrons in a small radiation phase, according to the theory of superradiance, the spectral power of the output radiation would further increase by a factor of $N^2$.

The study in this paper suggests several promising applications. For instance, a TEM equipped with a built-in coherent photon source could be useful for multi-dimensional pump-probe material studies. In addition, recently quantum phenomena are being observed in the interaction between keV electrons and a nano-grating[36,37]. Whether or not a nano-grating excited by a single keV electron could become an on-demand single-photon source for quantum optics is an interesting area for further investigation.

## AUTHOR DECLARATIONS

### Conflict of Interest

The authors declare no conflicts of interest.

## DATA AVAILABILITY

The data that support the findings of this study are available from the corresponding authors upon reasonable request.

## ACKNOWLEDGMENTS

This work is supported by the Ministry of Science and Technology, Taiwan, under Grants 110-2221-E007-103 and 108-2112-M-007-MY3, and the Swedish Foundation for Strategic Research under Project STP19-0081. Huang thanks helpful discussions with Joel England of SLAC, Andrzej Szczepkowicz of University of Wroclaw, and Levi Schachter of Technion-Israel Institute of Technology.